# Evolutionary game theory elucidates the role of glycolysis in glioma progression and invasion

David Basanta[1]*, Matthias Simon[2], Haralambos Hatzikirou[1] and Andreas Deutsch[1]

[1]Zentrum für Informationsdienste und Hochleistungsrechnen, Technische Universität Dresden, Dresden, Germany
[2]Neurochirurgische Universitätsklinik, Universität Bonn, Bonn, Germany

**Abstract**. Tumour progression has been described as a sequence of traits or phenotypes that cells have to acquire if the neoplasm is to become an invasive and malignant cancer. Although the genetic mutations that lead to these phenotypes are random, the process by which some of these mutations become successful and spread is influenced by the tumour microenvironment and the presence of other phenotypes. It is thus likely that some phenotypes that are essential in tumour progression will emerge in the tumour population only with the prior presence of other different phenotypes. In this paper we use evolutionary game theory to analyse the interactions between three different tumour cell phenotypes defined by autonomous growth, anaerobic glycolysis, and cancer cell invasion. The model allows to understand certain specific aspects of glioma progression such as the emergence of diffuse tumour cell invasion in low-grade tumours. We find that the invasive phenotype is more likely to evolve after the appearance of the glycolytic phenotype which would explain the ubiquitous presence of invasive growth in malignant tumours. The result suggests that therapies which increase the fitness cost of switching to anaerobic glycolysis might decrease the probability of the emergence of more invasive phenotypes.

# Introduction

Cancer cells display characteristic traits acquired in a step-wise manner during carcinogenesis. Some of these traits are autonomous growth, induction of neo-angiogenesis, and invasion and metastasis (Hanahan and Weinberg, 2000). Further characteristics of cancer cells include an altered glucose metabolism: tumour cells commonly switch to glycolysis for energy production (Warburg, 1930). FDG-PET (fluoro-D-deoxyglucose positron emission tomography) and MRS (magnetic resonance spectroscopy) imaging allows for an in-vivo analysis of glucose metabolism and lactate concentrations. The clinical use of PET imaging and MRS has largely confirmed the ubiquitous switch to glycolysis in many cancers including malignant gliomas (Herholz et al., 1992; Padma et al., 2003; McKnight, 2004). Glycolysis is less efficient than the citrate cycle in terms of energy, but allows for survival in hypoxic environments, i.e. when oxygen demands by the growing number of tumour cells are no longer met by the vascular supply of the tumour. In addition, cells with a glycolytic metabolism can change the pH of the local microenvironment to their advantage. Cells with a non-glycolytic metabolism will often undergo apoptosis or necrosis after prolonged exposition to acidic conditions. It has been suggested that tumour cells with a glycolytic metabolism evolve because of this fitness advantage over other cells (Gatenby et al., 2006).

---

*[1]Correspondence: Integrated Mathematical Oncology, H. Lee Moffitt Cancer Center & Research Institute, 33612, Tampa, Fl, USA. Email: david.basanta@moffitt.org

The prevalence of glycolysis in invasive tumours suggests that its presence could help the emergence of invasive phenotypes. Mathematical tools such as Game Theory (GT) can be used to study how the interplay between different phenotypes affects the outcome of tumour progression. GT models, first formalised by von Neumann and Morgernstern (von Neumann and Morgernstern, 1953), have a comparatively long tradition in the social and economic sciences. GT has also been successfully applied to the study of the evolutionary dynamics in nature (Nowak, 2006). More recently, GT has emerged as a tool in theoretical medicine. GT studies the interactions between entities called players where the fitness of each player depends on what the player decides to do (its strategy) as well as what the other players do. The players will obtain a fitness payoff as a result of these interactions. Maynard Smith (Maynard Smith, 1982) helped to establish evolutionary GT (EGT) as a tool for the study of equilibria in ecosystems. In EGT the strategies of the players, i.e. their phenotypes, are not the result of rational analysis but rather a behavior shaped through natural selection.

In this paper we introduce a EGT model in which we explore the hypothesis put forward by Gatenby and colleagues (Gatenby et al., 2006) that glycolysis precedes tumour invasion and place the results in the context of gliomas, tumours of the central nervous system derived from glial cells. We have previously used a cellular automaton model to describe the growth of gliomas (Hatzikirou et al., 2005; Hatzikirou and Deutsch, 2007). The model did not include any provisions for the various glioma malignancy grades and malignant progression. In contrast, the present paper specifically focuses on these latter aspects of cancer biology. We start at a stage in tumourigenesis with cells that no longer require external growth factors and ignore external growth inhibitory signals (autonomous growth). We postulate that additional genetic mutations will either result in an invasive phenotype, or a switch to a glycolytic metabolism. The results show that a motile/invasive phenotype is more likely to evolve with the presence of glycolytic cells.

## Materials and methods

We assume that all tumour cells are initially characterised by autonomous growth (AG phenotype). Cells can switch to anaerobic glycolysis for energy production (GLY phenotype), or become increasingly motile and invasive (INV phenotype). We make no assumptions as to what genetic changes are necessary for mutations to occur. The fitness of a cell with a given phenotype depends on its interaction with other cells that may have a different phenotype. Interactions between phenotypes are described in table 1. The base payoff, 1, represents the maximum fitness for a tumour cell under ideal circumstances in which it has to share nutrients and space with no other cell. The table is parametrised using variables: k, n and c. Variable k represents the fitness cost incurred by the switch to the less efficient glycolytic metabolism (GLY), n represents both: the loss of fitness for a non-glycolytic cell to live in an acid environment as well as the fitness gain for a glycolytic cell that increases the acidity of the microenvironment. Finally, c is the cost of motility incurred by cells with the INV phenotype resulting from the reduced proliferation rate of motile/invasive cells (Giese et al., 1996).

*Table 1. Payoff table that represents the change in fitness of a tumour cell with a given phenotype interacting with another cell. The three phenotypes in the game are autonomous growth (AG), invasive (INV) and glycolytic (GLY). The base payoff in a given interaction is equal to 1 and the cost of moving to another location with respect to the base payoff is c. The fitness cost of acidity is n and k is the fitness cost of having a less efficient glycolytic metabolism. The table should be read following the columns, thus the fitness change for an invasive cell interacting with an autonomous growth would be 1 − c.*

|     | AG            | INV             | GLY                       |
| --- | ------------- | --------------- | ------------------------- |
| AG  | $\frac{1}{2}$ | 1-c             | $\frac{1}{2} + n - k$     |
| INV | 1             | $1 - \frac{c}{2}$ | 1-k                     |
| GLY | $\frac{1}{2} - n$ | 1-c         | $\frac{1}{2} - k$         |

Table 1 should be read following the columns. For instance, the fitness payoff for an AG cell interacting with another AG, E(AG, AG), is 1/2 since AG cells have to share the available resources. When an AG cell meets an INV then the INV cell will leave for another location, obtaining the base payoff minus the cost of motility 1 − c, whereas the AG cell gets access to all the available resources and thus the base payoff. When an AG cell meets a GLY cell they both have to share the available resources. Furthermore, the AG cell loses fitness due to the acidification of the environment. GLY cells never get the full base payoff since their metabolism is less efficient.

# Results

We have adopted a standard EGT analysis to study the equilibria between the different phenotypes in two scenarios: (1) mutation(s) leading to invasive cells capable of motility in a tumour composed of autonomous growth cells (AG, INV) and (2) a tumour containing autonomous growth, motile and glycolytic cells (AG, INV, GLY).

**Scenario 1: AG and INV**

This scenario studies the situation in which a mutation can confer motility/invasiveness to tumour cells that already capable of autonomous growth i.e. the tumor is populated by AG cells, which can acquire the INV phenotype.

A population of AG cells is immune to invasion from a mutant INV phenotype if the fitness of two AG cells interacting together is greater than that of a mutant INV cells interacting with an AG cell. Alternatively, if the fitness of an AG cell playing with a second one is the same than that of an INV cell playing an AG cell, then AG cell might still be an immune to invasion if the payoff of an AG cell playing an INV cell is greater than that of an INV cell playing another INV cell. The payoff table shows that as long as motility represents a non negligible cost then the only thing required by a population of AG cells to be immune to invasion by an INV phenotype is that the fitness payoff of two AG cells is greater or equal than the fitness payoff for an INV cell interacting with an AG cell. That is, the AG phenotype is an immune to invasion only if $c \geq 1/2$. In many cases a polymorphism of INV and AG phenotypes will result. In these cases the fitness of the INV phenotype should be the same as that of the AG phenotype. If p is the proportion of invasive(INV) cells in the tumour, AG refers to the autonomous growth phenotype and W(P) the fitness of phenotype P:

$$W(I) = W(AG) \rightarrow$$
$$W(INV) = A(AG) \rightarrow$$
$$pE(INV,INV) + (1-p)E(INV,AG) = pE(AG,INV) + (1-p)E(AG,AG) \rightarrow$$
$$p(1-\frac{c}{2}) + (1-p)(1-c) = p + (1-p)\frac{1}{2} \rightarrow$$
$$p = \frac{1-2c}{1-c}$$

**Scenario 2: AG, GLY and INV**

Here we study a tumour in which all three phenotypes coexist. Assuming that p refers to the proportion of INV cells and p' to the proportion of AG cells, the fitness of each of the phenotypes can be calculated as:

$$W(INV) = p\frac{c}{2} + 1 - c$$
$$W(AG) = p(\frac{1}{2}+n) + p'n(\frac{1}{2}-n)$$
$$W(GLY) = \frac{1}{2}p + np' + (\frac{1}{2} - k)$$

In equilibrium the fitness of the three phenotypes is the same. The values of p and p' can be deduced from these equations. From these equivalencies:

$$p = 1 - \frac{k}{n}$$
$$p' = \frac{2kn + k - ck - cn}{2n^2}$$

Figure 1 shows that the proportion of INV cells in the tumour, p, changes when we alter the values of the cost of having the GLY phenotype (k) and the cost of a normal cell living next to a GLY cell (n). For low values of k and high values of n the INV cells displace the other phenotypes from the tumour. This means that conditions favouring anaerobic glycolysis also favour tumour invasion. One possible explanation for this phenomenon is that the presence of glycolytic cells that increase the acidity of the environment will indirectly and comparatively reduce the costs of motility.

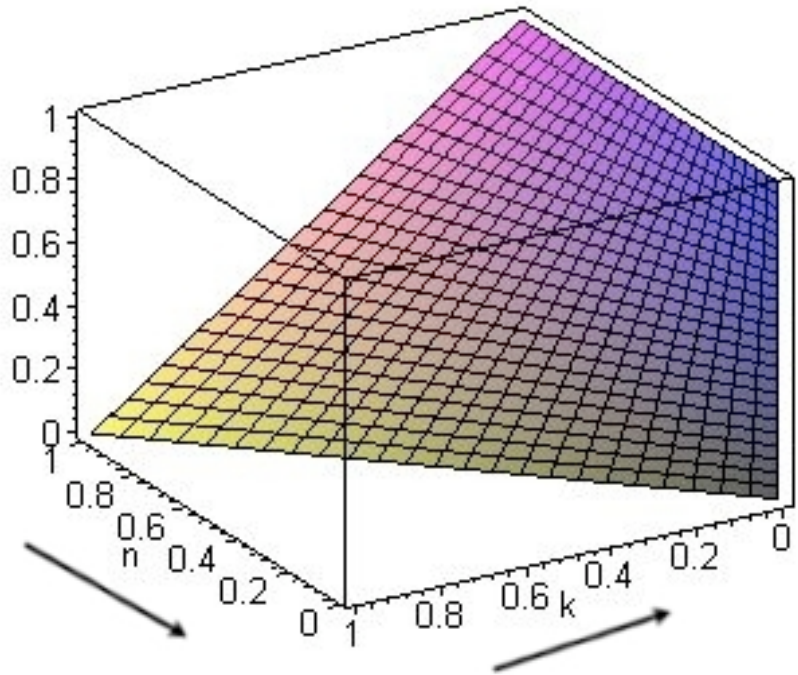

*Fig. 1. Proportion of invasive cells in a tumour with three phenotypes (autonomous growth, invasive and glycolytic). k is the cost in terms of fitness of adopting the glycolytic metabolism whereas n is the fitness cost of a normal cell when staying with a glycolytic cell.*

Figure 2 shows how the proportion of AG cells in the tumour population changes for variables k and n under four different scenarios for the cost of motility c. The proportion of AG cells increases as the cost of having a glycolytic metabolism increases or as the costs of living in an acid environment decreases thus reducing the proportion of both INV and GIY cells in the tumour. Interestingly, as the cost of motility increases the proportion of AG cells shrinks: higher costs of motility lead to more glycolytic cells that reduce the fitness of autonomous growth cells.

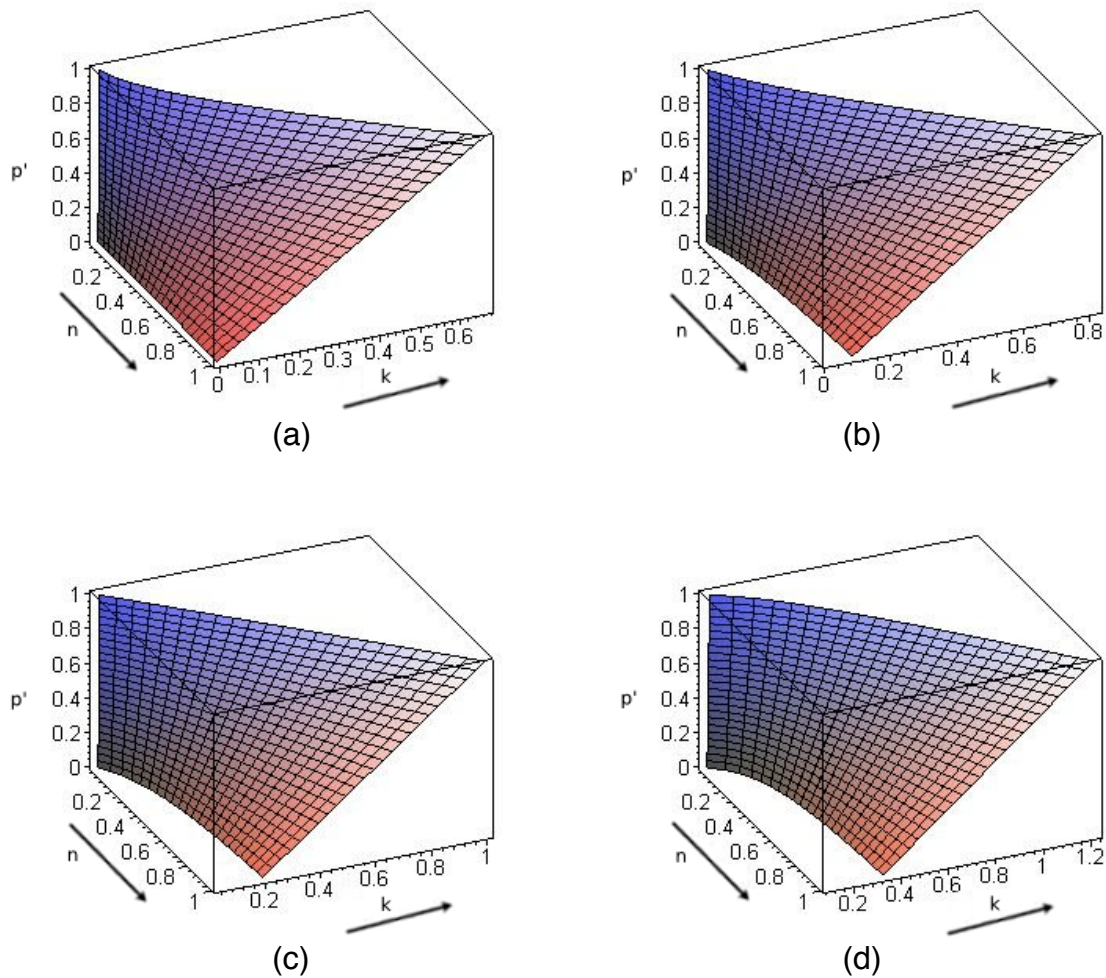

Fig. 2. The figures imply that a higher cost of motility leads towards tumours with a small proportion of AG cells. The figures (a), (b), (c) and (d) show the proportion of autonomous growth cells in a tumour in which the three phenotypes (autonomous growth, invasive and glycolitic) are possible. Each scenario is characterised by a different value of c, the cost of motility: (a) c = 0. (b) c = 1/4 (c) c = 1/2 (d) c = 3/4 .

# Discussion

Mathematical and computational models have been used to study tumour morphology and growth (Moreira and Deutsch, 2002; Anderson et al., 2006; Sanga et al., 2007) and therapy (Abbott and Michor, 2006). EGT has long been used to study the evolution of species but has only recently been applied to the analysis of the evolutionary aspects of cancer. Tomlinson and Bodmer (Tomlinson, 1997; Tomlinson and Bodmer, 1997) were the first to apply GT in cancer research. Their models study several cancer related problems such as angiogenesis and evasion of apoptosis. Subsequent research by Bach and colleagues extended this idea to interactions between three players (Bach et al., 2001) as well as the effect of spatial dynamics (Bach et al., 2003) on their angiogenesis game. Gatenby and Vincent adopted a game theory approach influenced by population dynamics to study tumour-host interactions in colorectal carcinogenesis (Gatenby and Vincent, 2003). More recently, Mansury and colleagues (Mansury et al., 2006) have employed a non-evolutionary GT approach to study how the interactions between proliferative and migratory phenotypes in a tumour affect a number of features of tumour growth dynamics.

To the best of our knowledge, ours is the first attempt to use evolutionary game theory to analyse the interplay of different tumor cell phenotypes with respect to tumor invasion. Our model makes certain predictions that can be compared to clinical observations. Virtually all gliomas including low-grade tumours display some degree of invasion. Tumour cell infiltration of several cerebral lobes is occasionally seen and our model provides a simple explanation for this observation, and the apparent difference to many other cancers. The model predicts that motile/invasive (INV) and autonomously growing cells (AG) will coexist as long as the fitness costs of motility are not too high when compared to the fitness increase obtained by moving to a new location. Glioma invasion in the brain is facilitated by the presence of white matter tracts that lower the costs of motility (Giese and Westphal, 1996). Relatively higher costs of motility are incurred by epithelial cancer cells, which have to transgress a formal barrier (the basal membrane separating the epithelial layer from the surrounding tissues), when compared to glioma cells. Furthermore, an invasive glioma cell will arrive in a surrounding very similar to the location from which it originated, i.e. in brain tissue. In contrast, an epithelial cancer cell leaves the context of other epithelial cells in order to grow among soft tissue cells and intracellular matrix molecules which may mean further fitness costs.

Malignant progression is common and more than 50% of low-grade tumours will eventually become malignant (Schmidt et al., 2003). In the model, the emergence of the glycolytic phenotype is the correlate of malignant progression. FDG-PET and SRS imaging has shown a close correlation between histological malignancy grades and anaerobic glycolysis in gliomas (Herholz et al.; 1992, Padma et al.; 2003, McKnight, 2004).

The results show that in a tumour populated by glycolytic cells, which in this model correlate with malignancy, invasive cells have a better chance of success (figure 1). Circumstances that favour the glycolytic phenotype also promote the emergence of the invasive phenotype. The proportion of invasive cells does not depend on the costs of motility, if all three phenotypes (AG, GLY, INV) are analysed in combination. This may reflect one of the fundamental tenets of malignant tumour progression. Malignant (glycolytic) tumours will also invariably, regardless of the specifics of the costs incurred, display a highly invasive phenotype. Invasion is a prerequisite for metastasis and, in most cases, characterises malignant tumours. Gatenby and co-workers have suggested a direct link between the glycolytic phenotype and tumour invasion and have provided experimental evidence to support this hypothesis (Gatenby et al., 2006).

Invasive growth is not as closely related to malignancy in gliomas. As pointed out above, in the absence of glycolytic cells, the costs of motility do influence the emergence of the invasive phenotype, possibly reflecting the very specific environment of the brain. Hence, the model provides an explanation for the frequently invasive growth of slowly growing (benign AG) low-grade gliomas. Invasive growth is usually not seen in most other cancers with the AG phenotype. Malignant glial tumours (glial tumors with a GLY phenotype) infiltrate the surrounding brain parenchyma, as predicted by the model. Importantly, glycolysis has been shown to support astrocytoma invasion in a tissue culture model (Beckner et al., 2005).

A mathematical model is necessarily a simplification of the complex situation found in a real tumour. Nevertheless, mathematical models provide theoretical frameworks in which to study qualitatively and quantitatively tumour progression (Gatenby and Maini, 2003; Anderson et al., 2006). Probably the most significant omission in our model is the lack of spatial considerations. Our previous research (Basanta et al., 2008) shows that game theoretical models of tumour invasion can have a similar predictive power than spatial ones based on Cellular Automata. Another possible future enhancement of the model would include changes in the way in which the fitness payoffs are considered. At the moment the costs of motility, acidity and glycolysis are considered to be constant. A more realistic approach would be to express these costs in terms of functions that could cover facts such as the increase in the costs of motility as the tumour grows leaving less space in which to move. Such enhancement would undoubtedly enrich the model but also make the analysis more difficult. Further research may also include the use of a larger number of phenotypes in the model, which are defined by molecular genetic aberrations (or sets thereof ), in order to allow for predictions of other important milestones in carcinogenesis.

Our model explains in mathematical terms specific aspects of glioma growth i.e. invasion already in slowly growing tumors, and puts them in a more general context. In addition, mathematical modelling of tumour progression may enhance our understanding of cancer and provide new insights. Our findings suggest that strategies such as improving tissue oxygenation which increase the relative fitness costs of switching to anaerobic glycolysis may prevent invasion and metastasis.

# Acknowledgements

We would like to acknowledge the help from Michael Kücken from Technische Universität Dresden. The research was supported in part by funds from the EU Marie Curie Network "Modeling, Mathematical Methods and Computer Simulation of Tumour Growth and Therapy" (EU-RTD IST-2001-38923). We also acknowledge the support provided by the systems biology network HepatoSys of the German Ministry for Education and Research through grant 0313082C.

# References

Abbott, L. H. and Michor, F. (2006). Mathematical models of targeted cancer therapy. Brit. Jour. Cancer, 95:1136–1141.

Anderson, A., Weaver, A., Cummings, P., and Quaranta, V. (2006). Tumor morphology and phenotypic evolution driven by selective pressure from the microenvironment. Cel l, 127:905–915.

Bach, L. A., Bentzen, S. M., Alsner, J., and Christiansen, F. B. (2001). An evolutionary game model of tumour cell interactions: possible relevance to gene therapy. Eur. Jour. Cancer, 37:2116–2120.

Bach, L. A., Sumpter, D. J. T., Alsner, J., and Loeschke, V. (2003). Spatial evolutionary games of interaction among generic cancer cells. J. Theor. Med., 5(1):47–58.

Basanta, D., Hatzikirou, H., and Deutsch, A. (2008). Studying the emergence of invasiveness in tumours using game theory. To appear in Eur. Phys. J. B.

Beckner, M., Gobbel, G., Abounader, R., Burovic, F., Agostino, N., Laterra, J., and Pollack, I. (2005). Glycolytic glioma cells with active glycogen synthase are sensitive to PTEN and inhibitors of PI3K and gluconeogenesis. Lab Invest., 85(12):1457–70.

Gatenby, R., Gawlinski, E., Gmitro, A., Kaylor, B., and Gillies, R. (2006). Acid-mediated tumor invasion: a multidisciplinary study. Cancer Res., 66(10):5216–23.

Gatenby, R. and Maini, P. (2003). Cancer summed up. Nature, 421:321.

Gatenby, R. and Vincent, T. (2003). An evolutionary model of carcinogenesis. Cancer Res., 63:6212–6220.

Giese, A., Loo, M., Tran, N., Haskett, S. W., and Berens, M. E. (1996). Dichotomy of astrocytoma migration and proliferation. Int. J. Cancer, 67:275–282.

Giese, A. and Westphal, M. (1996). Glioma invasion in the central nervous system. Neurosurgery, 39(2):235–50.

Hanahan, D. and Weinberg, R. (2000). The hallmarks of cancer. Cell, 100:57–70.

Hatzikirou, H., and Deutsch, A. (2007). Cellular Automata as microscopic models of cell migration in heterogeneous environments. Curr. Top. Dev. Biol. 81.

Hatzikirou, H., Deutsch, A., Schaller, C., Simon, M., and Swanson, K. (2005). Mathematical modelling of glioblastoma development. Math. Mod. Meth. Appl. Sci., pages 1779–1794.

Herholz, K., Heindel, W., Luvten, P., den Hollander, J., Pietrzyk, U., Voges, J., Kugel, H., Friedmann, G., and Heiss, W. (1992). In vivo imaging of glucose consumption and lactate concentration in human gliomas. Ann Neurol., 31(3):319–237.

Mansury, Y., Diggory, M., and Deisboeck, T. S. (2006). Evolutionary game theory in an agent based brain tumor model: exploring the genoype phenotype link. J. Theor. Biol., 238:146–156.

Maynard Smith, J. (1982). Evolution and the theory of games. Cambridge University Press, Cambridge.

McKnight, T. (2004). Proton magnetic resonance spectroscopic evaluation of brain tumor metabolism. Semin Oncol., 31(5):605–17.

Moreira, J. and Deutsch, A. (2002). Cellular automaton models of tumor development: a critical review. Advances in Complex Systems, 2-3:247–267.

Nowak, M. (2006). Evolutionary dynamics. Belknap.

Padma, M., Said, S., Jacobs, M., Hwang, D., Dunigan, K., Satter, M., Christian, B., Ruppert, J., Bernstein, T., Kraus, G., and Mantil, J. (2003). Prediction of pathology and survival by FDG PET in gliomas. J Neurooncol., 64(3):227–37.

Sanga, S., Frieboes, H. B., Zheng, X., Gatenby, R., Bearer, E. L., and Cristini, V. (2007). Predictive oncology: a review of multidisciplinary, multi- scale in silico modeling linking phenotype, morphology and growth. Neuroimage, 37 Suppl 1:S120–34.

Schmidt, M., Berger, M., Lamborn, K., Aldape, K., McDer- mott, M., Prados, M., and Chang, S. (2003). Repeated operations for infiltrative low-grade gliomas without intervening therapy. J Neurosurg., 98(6):1165–9.

Tomlinson, I. P. M. (1997). Game theory models of interactions between tumour cells. Eur. Jour. Cancer, Vol 33, N9, pp. 1495-1500.

Tomlinson, I. P. M. and Bodmer, W. F. (1997). Modelling the consequences of interactions between tumour cells. Brit. Jour. Cancer, 75(2):157–60.

von Neumann, J. and Morgernstern, O. (1953). Theory of games and economic behaviour. Princeton University Press, Princeton, NJ.

Warburg, O. (1930). The metabolism of tumors (English translation by F. Dickens). London: Constable.